# "Growing Evanescent Envelopes" and Anomalous Tunneling in Cascaded Sets of Frequency-Selective Surfaces in Their Stop Bands

*Andrea Alù*
*Dept. of Applied Electronics, University of Roma Tre, Rome, Italy.*
*Nader Engheta*[*]
*Dept. of Electrical and Systems Engineering, University of Pennsylvania, Philadelphia, PA, USA.*

*Abstract*: The presence of wave tunneling and the "growing evanescent envelope" for field distributions in suitably designed, periodically layered stacks of frequency selective surfaces (FSS) is discussed in this paper. It is known that a slab of double-negative (DNG) or left-handed metamaterial sandwiched between two vacuum half-spaces may lead to "growing" evanescent plane waves, which may be used to restore sub-wavelength information in Pendry's "perfect" lens [*Phys. Rev. Lett.*, 85, 3966, (2000)]. In this paper, it is shown that a completely different setup allows an analogous buildup of evanescently modulated waves. In particular, it is shown how an interface resonance phenomenon similar to the one present at the interface between metamaterials with oppositely signed constitutive parameters may be induced by a proper choice of the periodicities of the FSS stacks and the geometrical properties of these surfaces. The analysis is performed through an equivalent transmission-line approach, and some physical insights into this phenomenon are presented. Salient features, such as the complete wave tunneling through the pair of cascaded FSS, each operating at its bandgap, are presented and discussed.

**INTRODUCTION**
In his paper describing the idea of "perfect lens" using the slab of properly selected double-negative (DNG) media, i.e., a medium in which the real permittivity and permeability are simultaneously negative (also known as "left-handed" or "negative index" media), Pendry theoretically showed that inside such a metamaterial slab the evanescent wave has a "growing exponential", which contributes to the higher resolution of the image [1]. This idea has generated a lot of interests in the field of metamaterials, and many groups worldwide have been recently studying this issue of reconstruction of the evanescent waves and its sub-wavelength focusing. A natural question may arise: "May the growing exponential happen in other classes of media or in properly designed structures, perhaps arguably more easily realizable than DNG media?" Some ideas have been already reported in the literature [2], [3]. We have been interested to explore the answers to this question and to find other alternatives. Here, we show that a somewhat analogous phenomenon in the form of growth of evanescently modulated waves may occur inside suitably designed periodic arrangements of stacks of frequency-selective surfaces (FSS) within conventional (i.e., double-positive (DPS)) media.

Frequency Selective Surfaces (FSS), which are constructed as planar 2-dimensional periodic arrays of metallic elements with specific geometrical shapes, or as periodic apertures in a metallic screen, have been widely studied and known for many years [4]. The transmission and reflection coefficients for these surfaces are dependent on the frequency of operation (hence their names) and may also depend on the polarization and angle of incidence. In particular, the possibility of having frequency bands at which a given FSS is completely opaque (stop-bands) and others at which the same surface allows wave transmission has attracted the attention of scientists and designers.

In this work, we show how by periodically stacking two sets of FSS with "dual" behaviors (e.g., one stack of FSS with holes in the metallic plate, and the other a stack of FSS with metallic patches over the surface of a very thin dielectric holder*)* one may achieve growing field distributions towards the interface between these two stacks, and may have complete tunneling of wave through this pair, even though each stack alone would be "below cut-off" (i.e., in its stop band), similar to what we have obtained for an incident plane wave tunneling through a pair of properly chosen single-negative (SNG) bilayers [5].

---

[*] To whom correspondence should be addressed, engheta@ee.upenn.edu

**THEORETICAL ANALYSIS**

It is well known that an equivalent circuit model representation may be used for plane wave propagation and for its interaction with FSS [4]. In this transmission-line analogy, under a time-harmonic $e^{-i\omega t}$ excitation, an FSS may be represented as a shunt load, either inductive or capacitive depending on its geometrical parameters, in a transmission line (TL) segment representing the surrounding space. A cascade of FSS with dielectric materials filling the space between them (shown in Fig. 1, top part) would therefore be modeled as a series of TL cells periodically loaded by shunt elements, as in Fig. 1, bottom part. The distance between every FSS would correspond to the distance between the loads, and the "host" TL parameters $Z_c$ and $k_{TL}$ depend on the angle of incidence, the frequency and the polarization of the incident plane wave. The values of the shunt admittances representing the FSS depend directly on its geometrical parameters, in addition to the frequency, polarization, and angle of incidence of the incoming wave. Such parameters have been extensively studied and reported in the literature (see e.g., [4]) for several types of FSS. Obviously, in the limit of no loss, the values of the shunt admittances of FSS are assumed to be purely imaginary.

Referring to Fig. 1, therefore, the length of every single cell $d$ equals the distance between the two neighboring FSS and every TL unit cell has been chosen to have its shunt load in the middle (at $d/2$) to provide symmetry for the ease of analysis. The shunt admittance $Y = -iB$ represents the lossless FSS.

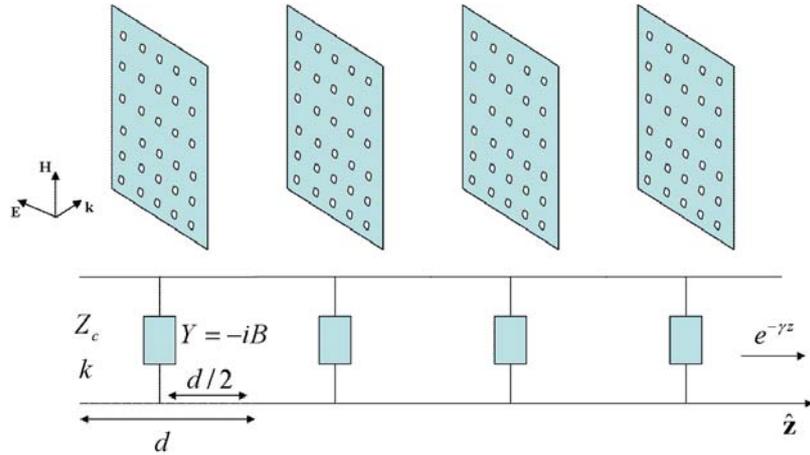

*Fig. 1 – A cascade of equally spaced FSS (top) and its circuit model representation (bottom)*

The problem of plane wave interaction with such a system, therefore, is equivalent to finding the solution for the voltage and current propagation along such a periodically loaded TL. The solutions can be determined from the well known Bloch equation (see e.g., [6]):

$$\cosh \gamma d = \cos k_{TL} d - \frac{BZ_c}{2} \sin k_{TL} d \ , \qquad (1)$$

where $\gamma$ is the Bloch wave number, which describes the $e^{-\gamma z}$-exponential dependence of the propagation/attenuation along $z$. A purely imaginary $\gamma$ corresponds to a propagating Bloch mode, whereas a positive real $\gamma$ corresponds to a decaying evanescent wave, representing the stop-band of this periodic structure. In the present work, we show how two different cascades of FSS (one with inductive FSS and the other with capacitive FSS), when properly designed, may exhibit "growing evanescently modulated" field distributions and complete tunneling of incident wave through this pair, although each cascade alone operates in its stop-band.

In our previous work [5] for a pair of only-epsilon-negative (ENG) and only-mu-negative (MNG) layers (or for a pair of DPS-DNG layers), we have derived the conditions for the transparency, resonance and complete tunneling of a wave incident on the pair, which depended on the characteristic impedances of the two metamaterials and their electrical lengths. In the case here, analogous conditions may be derived for wave transparency and interface resonance for the above-mentioned pair of cascades of dual FSS.

The Bloch impedance, which represents the ratio of voltage and current at the input section of every cell for an infinite periodically and homogeneously loaded TL, may be written as:

$$Z_B = \pm i Z_c \frac{\frac{BZ_c}{2}\cos k_{TL}d + \sin k_{TL}d - \frac{BZ_c}{2}}{\sqrt{\left(\cos k_{TL}d - \frac{BZ_c}{2}\sin k_{TL}d\right)^2 - 1}}. \quad (2)$$

Not surprisingly, by comparing (1) and (2) it can be verified that a real-valued $Z_B$ is possible only in the pass-bands, and conversely $Z_B$ can be imaginary in the stop-bands, when no real power is transferred. The sign should be positive when we are in the pass-bands (and $Z_B$ is real), whereas it may be positive or negative when $Z_B$ is imaginary, depending on the capacitive or inductive behavior of the evanescent decay along the TL.

Let us consider now the case of two different periodically loaded TL segments, with Bloch impedances $Z_{B1}$ and $Z_{B2}$ and electrical Bloch lengths $g_1 = N_1\gamma_1 d_1$ and $g_2 = N_2\gamma_2 d_2$, where $N_1$ and $N_2$ are the numbers of unit cells in the first and second cascades, stacked one after the other as in Figure 2a. We are not limiting ourselves to be only in the stop-bands or only in the pass-bands, i.e., the Bloch parameters may be in general complex. The input impedance at the connection between the two segments looking to the right may be found by the conventional TL formula:

$$Z_{in}(z = N_1 d_1) = Z_{B2} \frac{Z_c + Z_{B2} \tanh g_2}{Z_{B2} + Z_c \tanh g_2}. \quad (3)$$

The input impedance at the input of the whole system becomes:

$$Z_{in}(z=0) = Z_{B1} \frac{Z_{in}(z = N_1 d_1) + Z_{B1} \tanh g_1}{Z_{B1} + Z_{in}(z = N_1 d_1)\tanh g_1}. \quad (4)$$

Requiring that the system is totally transparent to the incident wave implies that $Z_{in}(z=0) = Z_c$. By inspection, the necessary and sufficient conditions to this equation are given by the two alternative solutions:

$$\begin{cases} Z_{B1} = -Z_{B2} \\ g_1 = g_2 \end{cases} \quad \begin{cases} Z_{B1} = Z_{B2} \\ g_1 = -g_2 \end{cases}, \quad (5)$$

which are formally equivalent to those derived in [5] for metamaterial bilayers. For passive loads, the condition on the right clearly corresponds to Bloch segments in their pass-bands (with imaginary $\gamma$ and positive real $Z_B$), whereas the condition on the left requires the two segments to be both in the stop-band (with positive $\gamma$ and conjugate imaginary $Z_B$).

When the left condition is applied, if one FSS cascade in its stop-band has an inductive behavior (i.e., $Z_{B1} = -i|Z_{B1}|$) the other should have a capacitive behavior ($Z_{B2} = i|Z_{B2}|$). An "interface resonance" in this case may occur, analogous to what we have shown for the ENG-MNG bilayers [5], leading to the condition for having "growing exponential" modulated waves. Proper selection of electrical lengths for the two cascades satisfying the condition $N_1\gamma_1 d_1 = N_2\gamma_2 d_2$, moreover, leads to a complete tunneling of the evanescent Bloch waves through the overall system. In other words, if each of the two FSS cascades alone is illuminated by a plane wave, most of the incident energy is reflected back because it is designed to be at its band gap. However, by pairing the two FSS cascades that satisfy the left condition given in (2), an incident wave may tunnel through completely, resulting in no reflection from this pair. We also observe a "growing evanescent envelope" for the field distribution inside this pair. It is interesting to note that: a) the condition does not depend on the outer medium, i.e., the system is transparent independently on what surrounds it as long as the outside media on both sides are the same, and b) there is potentially no limitation on the total length of the whole system, but the condition is on the ratio between the two lengths. Of course, when realistic losses are considered a higher tunneling is expected for shorter segments, due to absorbing phenomena and to the fact that the condition on the characteristic impedances can no longer be satisfied exactly, however the phenomenon still remains conceptually unchanged. Both these peculiar characteristics had been shown in [5] for the ENG-MNG bilayer.

Fig. 2 presents the voltage distribution along the pair of FSS cascades shown in Fig. 2a. The TL representation of the system has also been depicted. Fig. 2b clearly shows that the voltage at the entrance face is completely recovered at the exit face, thus it has "tunneled through", and at every cell of the first cascade the Bloch mode experiences an exponentially "growing" envelope for the field distribution. In Fig. 2c, the first cascade is effectively removed and the only the second cascade is present. As can be seen, the voltage distribution in the remaining cascade exhibits a decaying envelope, and the reflected wave is non zero. A similar effect may be observed if we remove the second cascade and keep the first one only. By comparing Figs. 2b and 2c, we see the effect of pairing the two FSS cascades that results in the growing evanescent envelope at the interface.

The second condition in (5) corresponds to the propagating Bloch modes (i.e., when the FSS are both operating at their pass-band) with opposite behaviors; one having a forward propagation, while the other having a backward propagation. Also in this case, somehow analogous to the pair of DPS-DNG slabs [7], [8], a total transparency may be achieved with a proper design of the line parameters.

The phenomenon underlying this anomalous behavior, as also in the case of Pendry's lens, relies on the "interface resonance" present at the junction of the two FSS cascades. We should remember, in fact, that this analysis is valid in the steady-state regime, when the multiple reflections at the resonant interface have already built up. A detailed explanation of the circuit equivalence of such growing evanescent fields for Pendry's lens, has been given by us in [9].

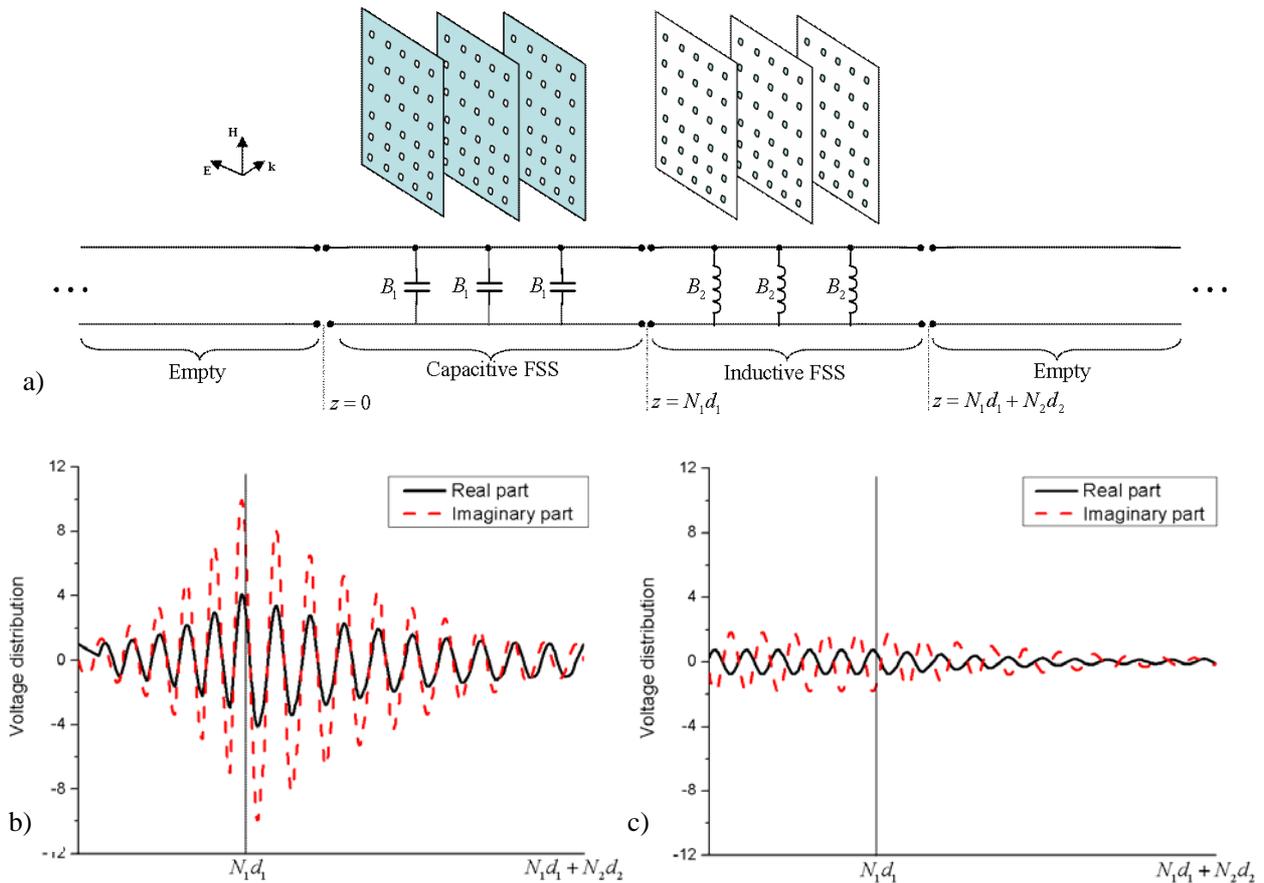

*Figure 2 – a) Stacked FSS cascades with "dual" behavior and their circuit model, e.g. one stack of FSS with holes in the metallic plate, and the other a stack of FSS with metallic patches over the surface of a very thin dielectric holder; b) Voltage distribution for two stacks satisfying the first of (2), $k_1 d_1 = 7\pi/4$, $k_2 d_2 = 0.49 + 2\pi$, $N_1 = 6$, $N_2 = 10$, $B_1 Z_c = 0.99$, $B_2 Z_c = -0.587$; c) The same, but $B_1 Z_c = 0$.*

**CONCLUDING REMARKS**

Here we have shown how even a simple collection of FSS with conjugate (dual) behavior, which is indeed easily constructible, may support the "growing" exponential modes that have attracted so much attention in recent years. In particular, it has been shown how a resonance phenomenon similar to the one present at the interface between ENG and MNG or DPS and DNG media may be induced by a proper choice of the spacing and the geometrical properties of the frequency selective surfaces. Interesting features, such as the presence of "growing" exponential envelopes and complete tunneling through the pair of stacked FSS, while each alone operating at its band gap, are presented. These results are considered a first step towards sub-wavelength imaging systems employing FSS structures.

**ACKNOWLEDGEMENTS**
This work is supported in part by the Fields and Waves Laboratory, Department of Electrical and Systems Engineering, University of Pennsylvania. Andrea Alù was supported by the scholarship "Isabella Sassi Bonadonna" from the Italian Electrical Association (AEI). Portions of these results were presented at the *2004 URSI International Symposium on Electromagnetic Theory, Pisa, Italy*, May 24-27, 2004.